# A comparison between two evaluations of neutron multiplicity distributions

J. P. Lestone
February 22$^{nd}$ 2008
Applied Physics Division, Los Alamos National Laboratory
Los Alamos, NM 87545, USA

**Abstract**

Within MCNP6, users can use multiplicity distributions for spontaneous and neutron-induced fission from an evaluation performed by Lestone (LA-UR-05-0288). This evaluation assumes the multiplicity distributions are Gaussian and adjusts the standard deviations to reproduce measured 2$^{nd}$ and 3$^{rd}$ factorial moments. In MCNPX, users have the option of using multiplicity distributions for spontaneous fission using an evaluation of the available experimental data by Santi and Miller (LA-UR-07-6229). The spontaneous fission multiplicity distributions and corresponding factorial moments from these two evaluations are compared in the present paper. The differences are minor or within experimental errors for all but the spontaneous fission of $^{238}$Pu. The $^{238}$Pu evaluations are based on data obtained in 1956. We recommend that the $^{238}$Pu multiplicity distribution be re-measured. Given the good agreement between the two evaluations, the choice of evaluation will make little difference for the modeling of spontaneous fission. However, we suggest that the evaluation in LA-UR-05-0288 be adopted as the MCNP default because it also includes recommended distributions for neutron induced fission.

I have been asked to comment on a debate related to the default neutron multiplicity distributions to be used in an upcoming merged version of MCNP6 and MCNPX. Within MCNP6, users can use multiplicity distributions for spontaneous and neutron-induced fission from an evaluation performed by Lestone [1]. In MCNPX, users have the option of using multiplicity distributions for spontaneous fission using an evaluation of the available experimental data by Santi and Miller [2]. The evaluation of Lestone assumes the multiplicity distributions are Gaussian. The standard deviations were adjusted to reproduce measured 2$^{nd}$ and 3$^{rd}$ factorial moments.

The mean number of fission neutrons emitted is a function of the fission fragment mass split. This dependence on the fragment mass split is predominantly caused by nuclear shell effects and exhibits the well known saw-tooth distributions. However, the physics behind the neutron multiplicity distributions is predominantly controlled by the distributions of fragment kinetic energies about the mean value for each mass split. For a given mass split, the fragment kinetic energy distribution closely follows a Gaussian [3]. To conserve total energy, the corresponding number of neutrons emitted from the fragments are anti-correlated with the kinetic energy of the fission fragments, and thus for each mass split, the neutron multiplicity distribution is nearly Gaussian. Integrating over all possible mass splits, it would be difficult for the underlining nuclear physics to conspire to give a final multiplicity distribution that differs significantly from a Gaussian. Model calculations that include the mass distribution, the dependence of the mean number of emitted neutrons on fragment mass split, the distribution of fragment kinetic energies for each mass split, and the experimentally known anti-correlations between fragment kinetic energy and neutron emission give neutron multiplicity distributions that are very close to Gaussians [3].

In general, experimental observables are more strongly related to the moments of the neutron multiplicity distribution than to the details of the distributions as a function of the number of emitted neutrons. Table I compares the 1st, 2nd, and 3rd factorial moments for the neutron distributions for spontaneous fission of U, Pu, Cm, and Cf isotopes as obtained using the evaluations of Lestone [1] and Santi and Miller [2]. The valuation of Santi and Miller includes an evaluation of the $^{238}$U spontaneous fission data. The evaluation of Lestone [1] does not include spontaneous fission of $^{238}$U, but does include neutron induced fission of U and Pu isotopes. A recommended Gaussian distribution for the $^{238}$U spontaneous fission neutron multiplicity distribution is presented in the present paper (see Table I, and Fig. 4).

Table I. 1st, 2nd, and 3rd factorial moments for the neutron multiplicity distributions for spontaneous fission of U, Pu, Cm, and Cf isotopes as obtained using the evaluations of Santi and Miller [2] and Lestone [1].

|   | $<\nu>$ | | $<\nu(\nu-1)>$ | | $<\nu(\nu-1)(\nu-2)>$ | |
|---|---|---|---|---|---|---|
|   | Ref [2] | Ref [1] | Ref [2] | Ref [1] | Ref [2] | Ref [1] |
| $^{238}$U* | 1.98±0.03 | 1.98 | 2.87±0.14 | 2.88 | 2.82±0.48 | 2.81 |
| $^{238}$Pu | 2.19±0.07 | 2.212 | 3.87 | 4.00 | 5.42 | 5.55 |
| $^{240}$Pu | 2.154±0.005 | 2.153 | 3.79±0.03 | 3.83 | 5.21±0.15 | 5.25 |
| $^{242}$Pu | 2.149±0.008 | 2.143 | 3.81±0.04 | 3.81 | 5.35 | 5.25 |
| $^{242}$Cm | 2.54±0.02 | 2.54 | 5.13±0.10 | 5.17 | 8.04±0.64 | 7.99 |
| $^{244}$Cm | 2.71±0.01 | 2.72 | 5.94±0.07 | 5.97 | 10.1±0.4 | 10.1 |
| $^{246}$Cm | 2.93±0.03 | 2.93 | 6.94±0.03 | 6.94 | 12.7±0.2 | 12.7 |
| $^{248}$Cm | 3.131±0.006 | 3.13 | 7.97±0.07 | 7.97 | 16.1±0.7 | 15.9 |
| $^{246}$Cf | 3.1±0.1 | 3.1 | 8.19 | 8.22 | 18.1 | 18.1 |
| $^{250}$Cf | 3.51±0.04 | 3.51 | 10.34 | 10.38 | 25.2 | 25.1 |
| $^{252}$Cf | 3.757±0.010 | 3.757 | 11.95±0.02 | 11.99 | 31.7±0.2 | 31.7 |

* The Gaussian distribution for $^{238}$U was not presented in ref [1]. The recommended standard deviation for the $^{238}$U spontaneous fission neutron multiplicity distribution is 1.135. This is 1.7% higher than the corresponding recommended value for $^{238}$U(n,f) [1].

The moments listed in Table I are displayed in figures 1, 2, and 3. The moments of the two evaluations are in excellent agreement. This is not surprising because the data sets used are almost the same. The evaluation of Santi and Miller used a larger data set including data measured since 2005. Lestone used evaluator judgment to exclude old data sets when more modern data were available. The U, Pu, Cm, and Cf spontaneous fission neutron multiplicity distributions of Santi and Miller [2] are compared to the recommended Gaussian distributions of Lestone [1] in figures 4-14. The agreement between the two recommended distributions is excellent for the U, Cm, and Cf isotopes. A minor difference exists for the $P_0$ and $P_1$ values for spontaneous fission of $^{240}$Pu and $^{242}$Pu. However, the agreement between the corresponding moments is excellent (see Table I). There is a difference between the two recommended $P_2$ and $P_3$ values for spontaneous fission of $^{238}$Pu. The $P_2$ of ref [2] is ~1.7 sigma higher and the $P_3$ of ref [2] is ~1.7 sigma lower than the corresponding recommended values of ref [1]. Notice that none of the other Pu, Cm or Cf distributions have a maximum in the distributions above 0.35 and all the other distributions are well characterized by Gaussians. The failure of the $^{238}$Pu distribution to follow the systematics of the other distributions suggests that the $^{238}$Pu spontaneous fission multiplicity distribution should be re-measured.

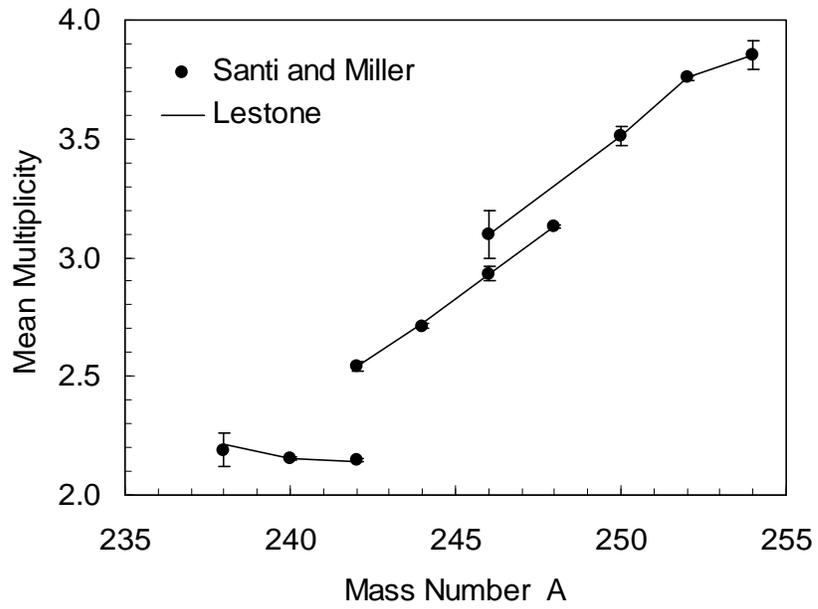

Fig. 1. 1st factorial moment for the neutron multiplicity distribution for spontaneous fission of Pu, Cm, and Cf isotopes as obtained using the evaluations of Santi and Miller [2] and Lestone [1].

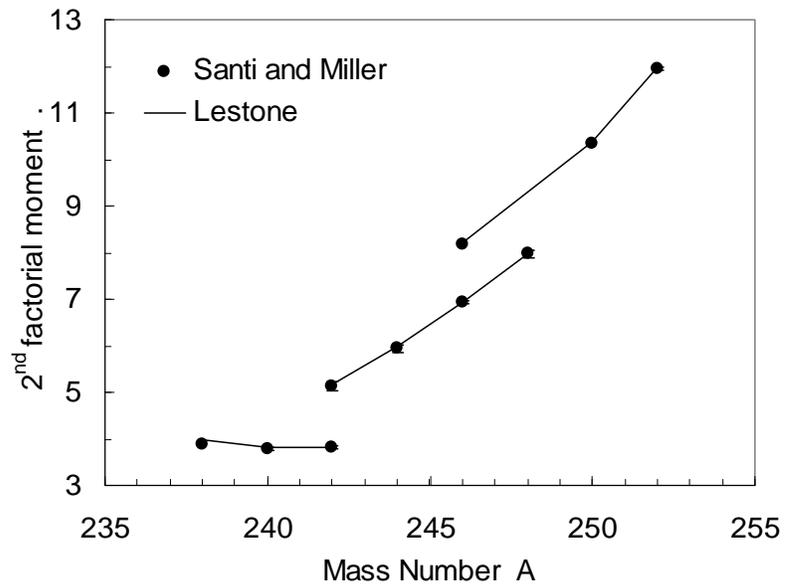

Fig. 2. As for Fig. 1 but for the 2nd factorial moment.

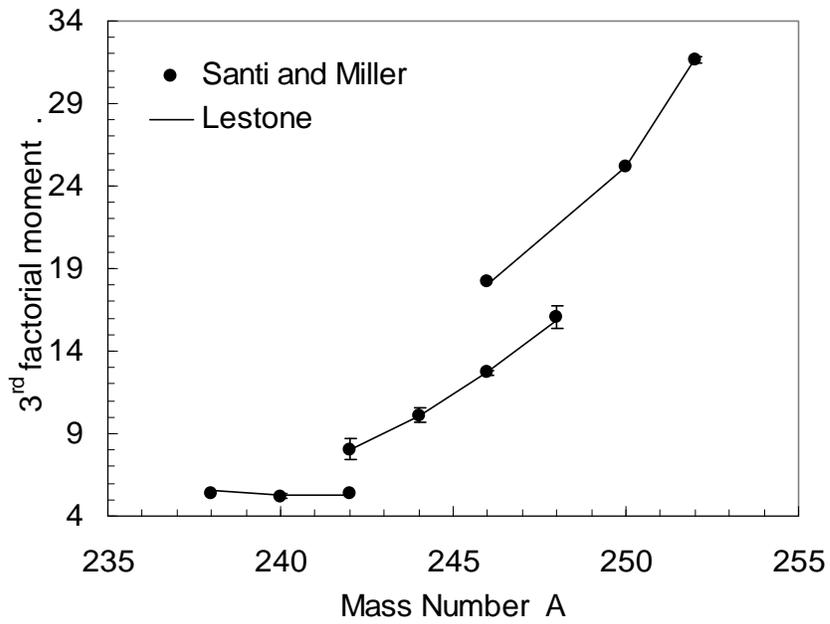

Fig. 3. As for Fig. 1 but for the 3$^{rd}$ factorial moment.

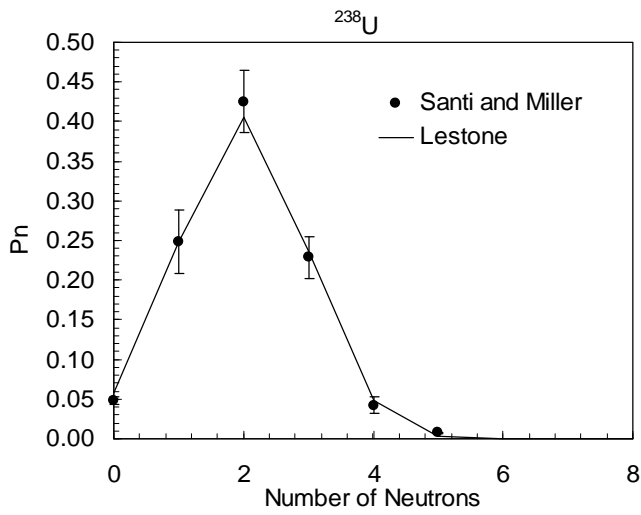

Fig. 4. $^{238}$U spontaneous fission neutron multiplicity distributions from two difference evaluations.

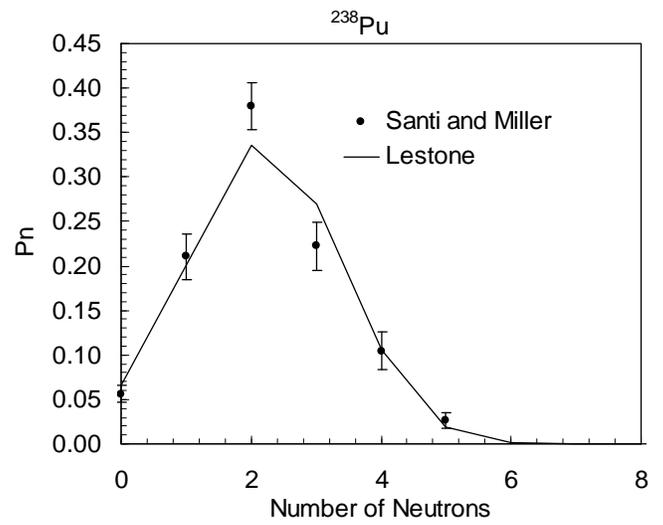

Fig. 5. As for Fig. 4 but for $^{238}$Pu.

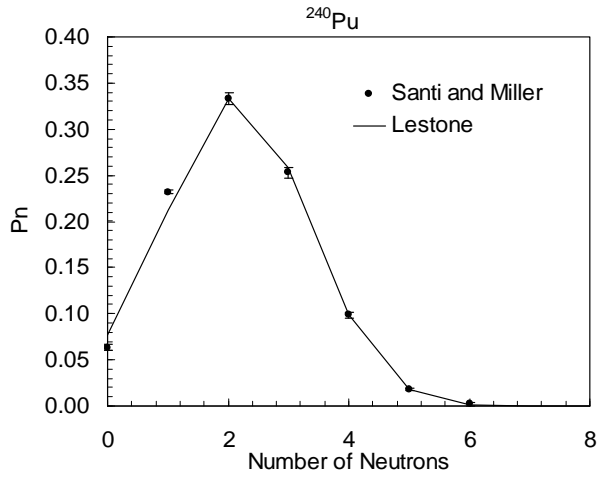

Fig. 6. As for Fig. 4 but for $^{240}$Pu.

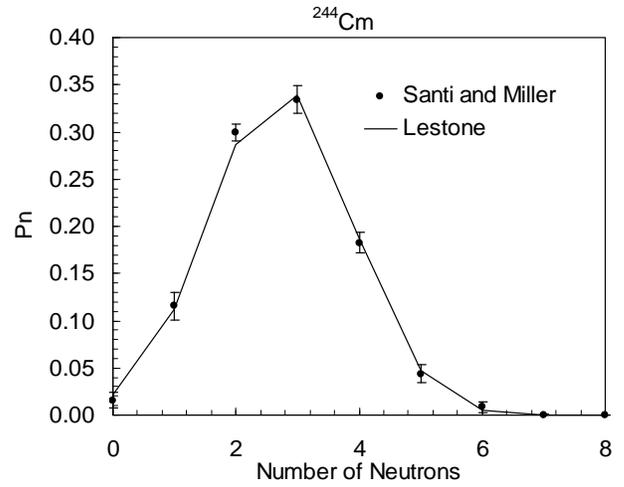

Fig. 9. As for Fig. 4 but for $^{244}$Cm.

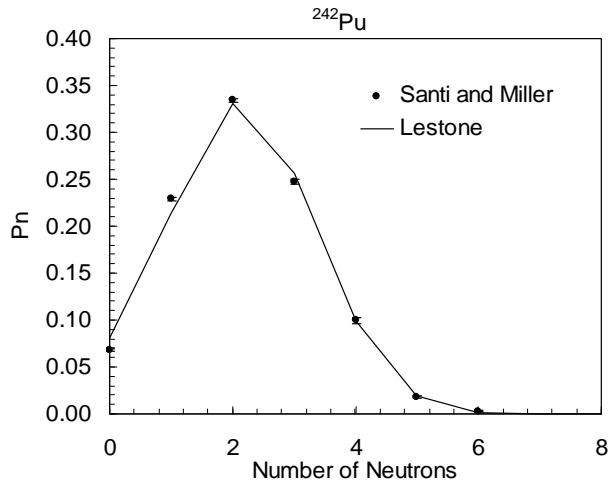

Fig. 7. As for Fig. 4 but for $^{242}$Pu.

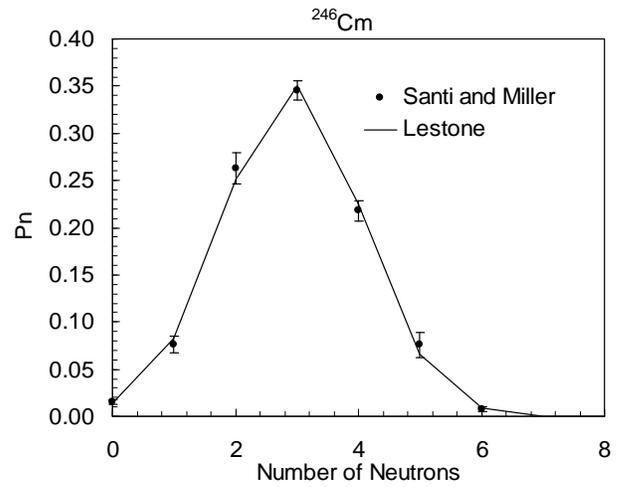

Fig. 10. As for Fig. 4 but for $^{246}$Cm.

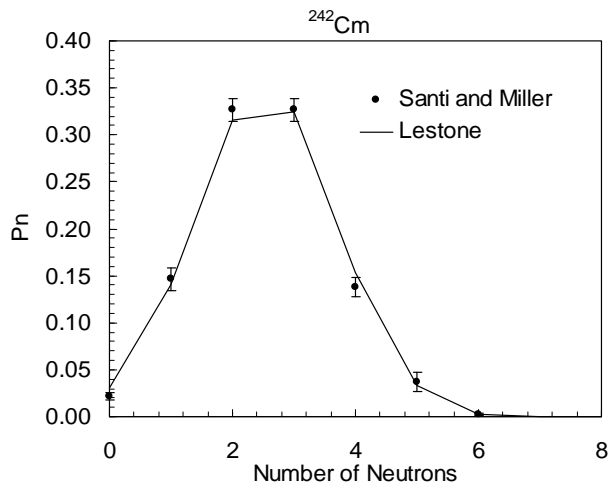

Fig. 8. As for Fig. 4 but for $^{242}$Cm.

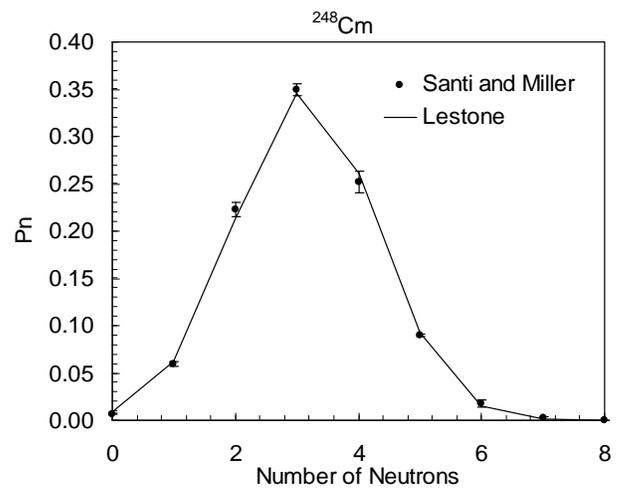

Fig. 11. As for Fig. 4 but for $^{248}$Cm.

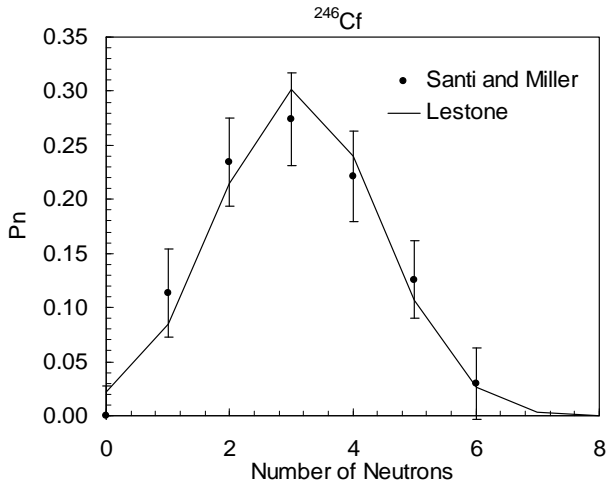
Fig. 12. As for Fig. 4 but for $^{246}$Cf.

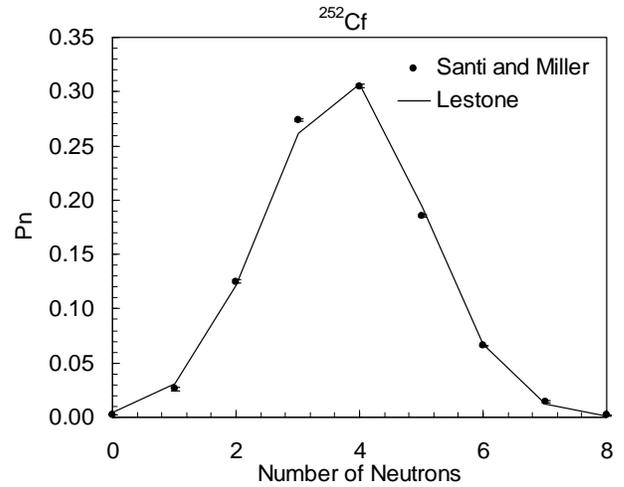
Fig. 14. As for Fig. 4 but for $^{252}$Cf.

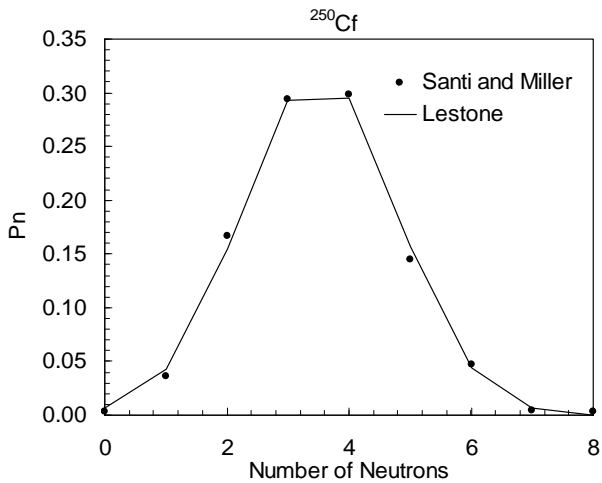
Fig. 13. As for Fig. 4 but for $^{250}$Cf.

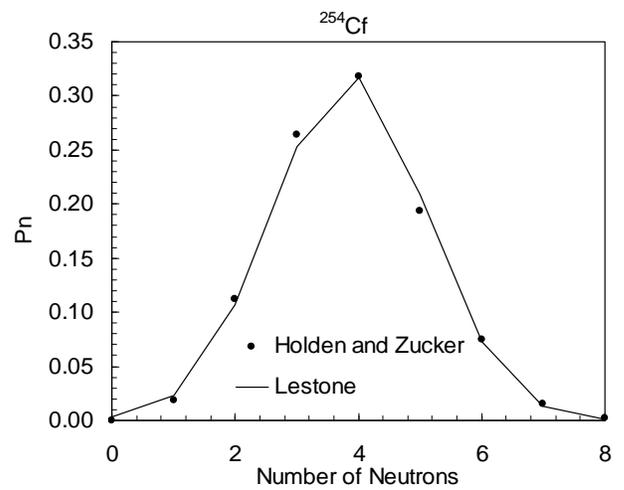
Fig. 15. As for Fig. 4 but for $^{254}$Cf.

To emphasis the point that fission neutron multiplicity distributions are well characterized by Gaussian distributions, the recommended Gaussian distributions for spontaneous fission of $^{254}$Cf and thermal neutron induced fission of U and Pu isotopes [1] are compared to the evaluation of Holden and Zucker [4] in figures 15-19. The key conclusion of ref [1] is that a constant width independent of incoming neutron energy is consistent with the available data for neutron induced fission up to an incoming neutron energy of 10 MeV.

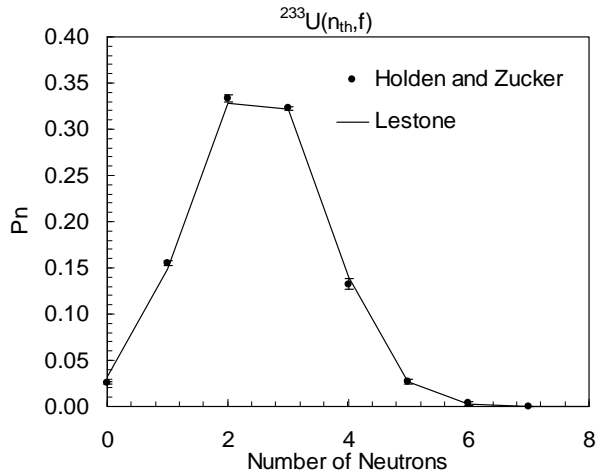

Fig. 16. As for Fig. 4 but for thermal neutron induced fission of $^{233}$U.

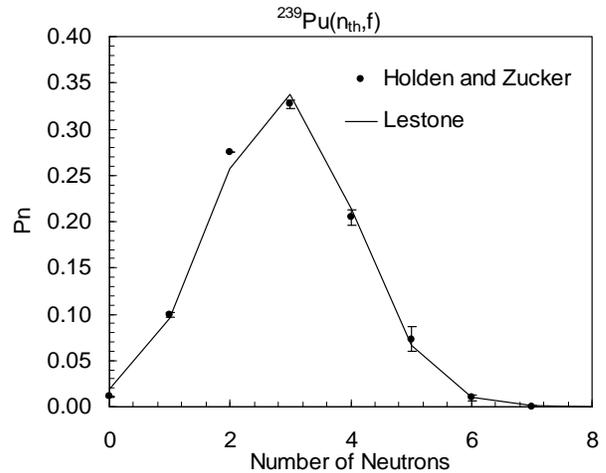

Fig. 18. As for Fig. 16 but for thermal neutron induced fission of $^{239}$Pu

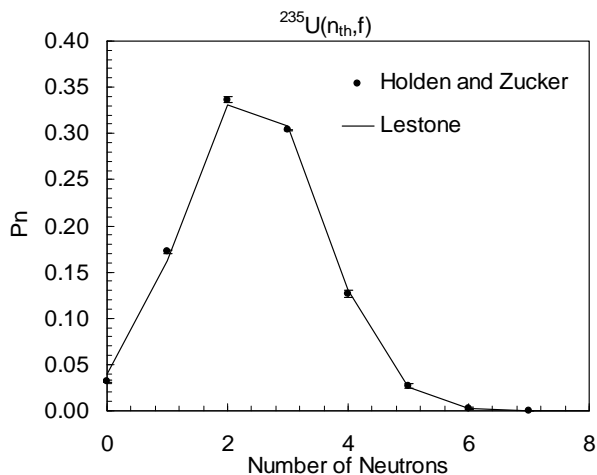

Fig. 17. As for Fig. 16 but for thermal neutron induced fission of $^{235}$U.

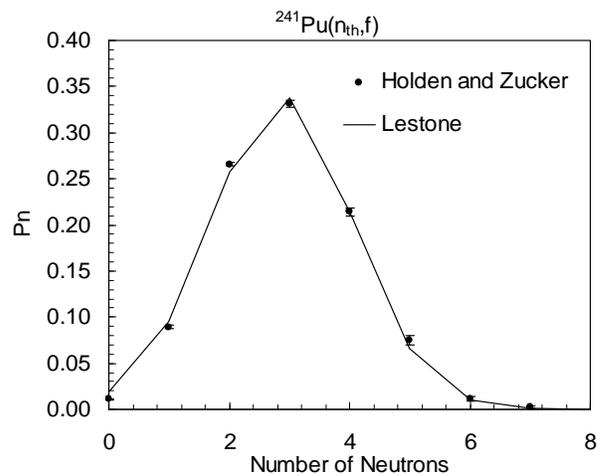

Fig. 19. As for Fig. 16 but for thermal neutron induced fission of $^{241}$Pu.

Given the good agreement between the two evaluations, the choice of evaluation will make little difference for the modeling of neutrons from spontaneous fission. However, we suggest that the evaluation in LA-UR-05-0288 be adopted as the default because it also includes recommended distributions for neutron induced fission.

**References**

[1] Energy and Isotope Dependence of Neutron Multiplicity Distributions, J. P. Lestone, Los Alamos National Laboratory Report (2005) LA-UR-05-0288.
[2] Reevaluation of Prompt Emission Multiplicity Distributions for Spontaneous Fission, P. Santi and M. Miller, Los Alamos National Laboratory Report (2007) LA-UR-07-6229.
[3] Subroutines to Simulate Fission Neutrons for Monte Carlo transport Codes, J. P. Lestone, Los Alamos National Laboratory Report (1999) LA-UR-99-5444.
[4] Prompt Neutron Emission Multiplicity Distributions and Average Values (Nubar) at 2200 m/s for Fissile Nuclides," Nucl. Sci. Eng., **98**, 174 (1988).